\documentclass[twocolumn,showpacs,preprintnumbers,amsmath,amssymb]{revtex4}
\usepackage{graphicx}% Include figure files
\usepackage{dcolumn}% Align table columns on decimal point
\usepackage{bm}% bold math

\begin{document}

\preprint{APS}

\title{
%Non-equilibrium
Green's function approach to transport through
 a gate-all-around Si nanowire under impurity scattering}
\author{Jung Hyun Oh}
\author{D. Ahn }
\email{dahn@uos.ac.kr}
\affiliation{Institute of Quantum Information Processing and
Systems, University of Seoul, Seoul 130-743, Korea}%
\author{Y. S. Yu}
\affiliation{Department of Information and Control Engineering,
Hankyong National University, Anseong 456-749, Korea}%
\author{S. W. Hwang}
\affiliation{School of Electrical Engineering, Korea University, Seoul 136-701, Korea}

\date{\today}% It is always \today, today,


\begin{abstract}
We investigate transport properties of gate-all-around Si nanowires
using  non-equilibrium Green's function technique.
By taking into account of the ionized impurity scattering we calculate Green's functions self-consistently and examine the effects of ionized impurity
 scattering on electron densities and currents.
For nano-scale Si wires, it is found that, due to the impurity scattering,
the local density of state profiles loose it's interference oscillations as well as is broaden and shifted.
In addition, the impurity scattering gives rise to
a different transconductance as functions of temperature and impurity scattering strength when compared with the transconductance without impurity scattering.

%We compare the variation of transconductance as a function of temperature
%for several impurity scattering strength.
\end{abstract}





\maketitle

\section{Introduction}

The study of ballistic electron transport in nano-devices has been an interesting field
of research.\cite{Thornton,Choi,Wees,DiCarlo}
Recently, a Si nanowire with a length comparable to the de Broglie wavelength
of carriers is realized by advanced nanofabrication technique.\cite{Nishiguchi,Cho}
The cross-sectional area of Si nanowires was designed to show well-separated transverse modes and electrons confined to the wire are expected to suffer from a minimal amount of impurity scattering.
These properties make the Si nanowires good candidates for the study of
ballistic quantum transport.
In addition, the potential distribution within the wire can be controllable by
a metallic gate around the wire. This provides additional degree-of freedom on currents through the device and one would expect that the basic transistor action is possible for a Si nanowire. As a result,
the gate-all-around Si wire may shed the light on
one-dimensional structures for future transistor applications.

It is desirable experimentally to make the Si wires as intrinsic as possible.
However, to populate the wires with carriers, it is necessary to define source and drain regions
where ionized dopants are placed.
These dopants scatter free carriers and the elastic impurity scattering cannot be avoided
in those regions.  Thus, in order to understand transport in the wires, 
a quantitative treatment of the ionized impurity scattering will be important.
Several theoretical works were done to investigate the effects of ionized-impurity
scattering on one-dimensional electron gas, and revealed their effects on the electronic structure.
Most of these studies were for  uniformly doped or remote-impurity systems\cite{Masek,Hu}
and 
adopted empirical models based on the so-called B\"{u}ttiker probes for simulating
the device.\cite{Venugopal,Wang}
The empirical methods are appealing due to relatively simple implementation but the methods often require parameters that need to be adjusted using more rigorous calculations or values from experiments.

In this work, we take into account the ionized impurity scattering
in simulating the gate-all-around nanowire using non-equilibrium
Green's function approach. 
By averaging the Green's function over impurity configurations and expanding the arising term perturbatively,
we treat the impurity scattering within a self-consistent Born approximation and
apply the formula to the Si nanowire as realized in Ref. \cite{Cho}.
Since the impurity-scattering strength is a single parameter for the system,
the method provides the first-principle approach to understand current-voltage characteristics and compare them with the experimental results.

%We also present expressions for local electron densities and currents with the averaged

%Green's functions.

\section{Calculation method}

\subsection{Hamiltonian}

\begin{figure}
\centering
\includegraphics[width=0.4\textwidth]{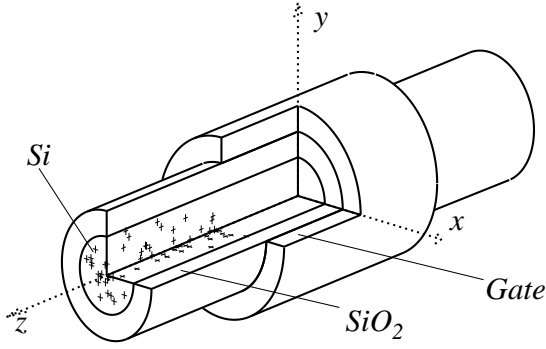}
\caption{\label{figSt}
We plot a schematic diagram of a cylindrical Si wire simulated in this work
which is oriented along the [001] direction.
The Si wire surrounded with the gate is assumed to be intrinsic and
separate the source and drain regions where ionized dopants are distributed.
}
\end{figure}

To see the effects of the impurity scattering clearly,
we consider a simple geometry of a quantum wire as in Fig. \ref{figSt}.
An infinitely-long cylindrical Si wire consists of intrinsic channel and heavily doped source and drain regions.
A metallic gate extended over a length of $L_G$
is rolled round the intrinsic region
and they are separated from each other by a SiO$_2$ layer
with a width $t_{\rm ox}$.
For simplicity, 
we assume that the Si wire is grown along the crystal
$[001]$-axis(chosen as the $z$ direction in the figure)
and the doping profile of $N_D({\bf r})$ in the source and drain regions is symmetric about the $z$-axis so that we can utilize the circular symmetry.

Then, electrons in the Si wire are governed by the effective-mass Hamiltonian which is given by
\begin{eqnarray}
\hat{H} &=& \int \hat{\psi}^*({\bf r})\Big\{   
-\frac{\hbar^2 }{2}\left(
 \frac{1}{m_x}\frac{d^2}{dx^2}
+\frac{1}{m_y}\frac{d^2}{dy^2}
+\frac{1}{m_z}\frac{d^2}{dz^2}\right)\nonumber\\
&+& U({\bf r})+U_{imp}({\bf r})
\Big\}\hat{\psi}({\bf r}) d{\rm r}.
\label{H}
\end{eqnarray}
Above Hamiltonian describes electrons in six different valleys depending on their
effective masses.  For instance, if $m_x= m_z = 0.19m$, transverse mass, and
$m_y=0.95m$, longitudinal mass of Si,
the Hamiltonian represents electrons in the $[010]$-valley, etc.
Here, $U({\bf r})$ is the macroscopic potential energy resulted from
both band discontinuity among the materials, and the Coulomb contribution from external charges.
The Coulomb part is determined by the Poisson's equation,
\begin{eqnarray}
-\nabla^2 U({\bf r}) = \frac{e^2}{\epsilon_{Si}}\{ N_D({\bf r})-n_{el}({\bf r})\}
\label{Poisson}
\end{eqnarray}
when we know the electron distribution $n_{el}(\bf r)$.
$U_{imp}({\bf r})$ describes the impurity potential energy from the ionized dopants.
In this work, we assume that the impurity potentials are short-ranged but 
still vary slowly in the atomic scale.
As a result, different valley modes are not coupled by the impurity potential and
can be solved independently.

Since the device has the circular symmetry, it is convenient to express the Hamiltonian
in terms of the basis diagonalizing the radial motion. We choose the basis satisfying
the following Schr\"{o}dinger equation,
\begin{eqnarray}
\left [-\frac{\hbar^2 }{2}\left( \frac{1}{m_x}\frac{d^2}{dx^2}\! +\!\frac{1}{m_y}\frac{d^2}{dy^2}\right )
\!+\!U_{B}(\vec{\rho}) \right]\mid\! \chi_{l}\rangle \!=\!\epsilon_{l}\!\mid\! \chi_l\rangle
\end{eqnarray}
where $\vec{\rho}$ is radial coordinates $(x,y)$ and
$U_L(\vec{\rho})=U(\vec{\rho},\pm\infty)$ is a potential energy at $z=\pm\infty$,
i.e., in the deep source and drain regions.
Then, we expand the field operator $\hat{\psi}({\bf r})$ as,
\begin{eqnarray}
\hat{\psi}({\bf r}) = \sum_{ml} \hat{b}_{lm} \chi_l(\vec{\rho}) \psi(z_m)
\label{basis}
\end{eqnarray}
where we discretize the longitudinal coordinates with a spacing of $a$ and
$\psi(z_m)$ is tight-binding basis at the $m-$th node ($z_m=m a,~ m=-\infty,..\infty$).

Using Eq. (\ref{basis}) and a finite difference approximation, one can express
the Hamiltonian of Eq. (\ref{H}) as,
\begin{eqnarray}
\hat{H} = \sum_{lml'm'}\hat{b}^\dagger_{lm} \Big[
{\rm\bf H}_{lm:l'm'}+\!{\rm\bf v}_{lm:l'm'}+ \nonumber\\
\langle\! \chi_l\!\mid\!U_{imp}(\vec{\rho},z_m) \!\mid\! \chi_{l'}\!\rangle ]\delta_{mm'}
\Big ] \hat{b}_{l'm'}.
\label{Hp}
\end{eqnarray}
Here, the first term describes motion along the longitudinal direction
for each transverse mode and it's elements are given by,
\begin{equation}
{\bf H}_{lm:l'm'}\!=\![\delta_{m,m'}(\epsilon_{lm}+2t_H)\!-\!t_H(
\delta_{m,m'+1}\!+\!\delta_{m,m'-1})]\delta_{ll'} \nonumber
\end{equation}
with $\epsilon_{ml}\!=\!\epsilon_l\!+\!\langle
\chi_l\!\mid\! U(\vec{\rho},z_m)\!-\!U_L(\vec{\rho})\!\mid\! \chi_{l}\rangle$
and the hopping energy of $t_H = \hbar^2/2m_z a^2$
(hereafter, we use bold characters to denote a matrix
displayed on the basis $\{\chi_l\psi_m\}$).
The ${\bf v}$ matrix in Eq. (\ref{Hp})  accounts for the deviated potential distribution
from that of deep source and drain regions. As a result, it gives rise to the hybridization
among transverse modes as,
\begin{eqnarray}
{\bf v}_{lm:l'm'}&\!=\!&[\langle\! \chi_l\!\mid\! U(\vec{\rho},z_m)\!-\!U_L(\vec{\rho})
\mid\!\chi_{l'}\!\rangle (1-\delta_{l,l'})\delta_{mm'}.
\end{eqnarray}
%and, therefore, give rise to the coupling between transverse modes.
The last term in Eq. (\ref{Hp})  is a contribution from the impurity potential.

\subsection{ Impurity-averaged Green's function}

Now we formulate non-equilibrium Green's functions for the Hamiltonian of Eq. (\ref{Hp}).
In order to take into account the impurity scattering, we consider a number of impurity configurations
rather than a particular distribution, and average the Green's functions over the configurations.
For this we adopt the Schwinger-Keldysh technique.\cite{Kamenev}
According to the scheme, the impurity average gives rise to the quadratic interaction in the action,
and we expand it perturbatively to obtain the one-particle irreducible self-energy ${\bf\Sigma}^{imp}$.
Here, we restrict our attention to the first order diagram  and treat it  self-consistently,
which is referred to as the self-consistent Born approximation.\cite{Masek,Hu,Camblong,Lake}

The impurity-averaged Green's function ${\bf G}$ can be obtained through the Dyson's equation,
\begin{eqnarray}
{\bf G}(E) = {\bf g}(E)+{\bf g}(E){\bf \Sigma}^{imp}(E){\bf G}(E),
\label{Dyson}
\end{eqnarray}
where ${\bf g}$ is the impurity-free Green's function (in fact, the bold characters in this case
represent enlarged matrices for taking into account the Keldysh space. However, we keep the notation in
the meanwhile because it recovers an original size when we specify it's components explicitly in the Keldysh
space).
The corresponding self-energy from the impurity scattering depends on
it's Green's functions again through
the relation,
\begin{eqnarray}
{\bf\Sigma}^{imp}_{lm:l'm'}(E) =\sum_{l_1m_1 l_2m_2} S_{lml'm':l_1m_1l_2m_2} {\bf G}_{l_1m:l_2m}(E)
\label{ImpSigma}
\end{eqnarray}
with
\begin{eqnarray}
S_{lml'm':l_1m_1l_2m_2} &=& 
\frac{1}{2}
\Big\langle \langle \chi_l    \psi_m\!\mid\!  U_{imp}({\bf r}) \mid\chi_{l_1}\psi_{m_1}\rangle
\nonumber\\
&&\langle \chi_{l_2}\psi_{m_2}\!\mid\! U_{imp}({\bf r}')\!\mid\!\chi_{l'}\psi_{m'}\rangle\Big\rangle_{av}.
\end{eqnarray}
Here, $\langle \cdots\rangle_{av}$ denotes a configuration average.
We model fluctuating impurity potentials with a $\delta$-correlated function considering
the short-ranged form;
\begin{eqnarray}
\langle U_{imp}({\bf r}) U_{imp}({\bf r}')\rangle_{av} =
n_D({\bf r}) u_0^2~ l_s^3~ \delta({\bf r}-{\bf r}').
\end{eqnarray}
Here, $n_D({\bf r})=N_D({\bf r})/N_0$ is a normalized doping profile with respect to
the atomic density $N_0$ of Si.  And the impurity potential strength is
expressed with the impurity potential amplitude of $u_0$ and a screening length  $l_s =4$\AA,
which is approximately equal to the Tomas-Fermi screening length in the bulk Si at  carrier
density of $1\times 10^{20}/{\rm cm}^3$.
Accordingly, the expansion coefficient in Eq. (\ref{ImpSigma}) becomes,
\begin{eqnarray}
S_{lml'm':l_1m_1l_2m_2} =\frac{u_0^2 ~l_s^3}{2a} \delta_{mm'}\delta_{m_1m_2}\delta_{mm_1} \nonumber\\
\langle \chi_l\!\mid\! \chi_{l_1}(\rho) n_D(\rho,z_m)
\chi_{l_2}^*(\rho)\!\mid\!\chi_{l'}\rangle.
\label{Overlap}
\end{eqnarray}
It is noted that the short-ranged potential is diagonal for longitudinal basis $\{\psi_m\}$ but not for transverse modes $\{\chi_l\}$.
This means that transverse modes are mixed to each other through the impurity scattering.

For a given ${\bf\Sigma}^{imp}$, in order to solve the Dyson equation of Eq. (\ref{Dyson}),
we should take care of open-boundaries in our problem,
i.e., the infinite number of nodes along the longitudinal direction $(m=-\infty,...,\infty)$.
For this, we follow the conventional approach where the device is partitioned into the system being
in non-equilibrium and reservoirs.\cite{Datta}
Since the source and drain regions are extended semi-infinitely,
we confine our attention to the portion of the system near the gate
where physical properties are thought to be deviated from those of 
deep source
and drain regions.
We designate the portion by longitudinal indices $m=(0,1,...,M-1)$.
Thus, nodes for $m <0$ ($m\geq M$) represent the source (the drain) being in equilibrium 
with the chemical potential $\mu_S$ ($\mu_D$).

In the source and drain reservoirs, we assume that the self-energy ${\bf\Sigma}^{imp}$ is
independent of longitudinal coordinates $m$ because
they are sufficiently far from the gate region where the potential distribution is uneven.
Within this assumption, the Schr\"odinger equation is easily solved
and  equilibrium Green's functions ${\bf G}(E)$ with corresponding self-energies
are calculated straightforwardly.  In the Appendix, we illustrate their simple expressions.

Now, we focus on the device region, i.e., nodes ranging $0\leq m<M$
where one expects a non-equilibrium situation for different
chemical potentials of $\mu_S$ and $\mu_D$.
The Green's functions are obtained by truncating the matrix equation
of Eq. (\ref{Dyson}) within longitudinal indices of $0\leq m<M$.
Instead, the truncation 
introduces an additional self-energy
 $\tilde{\bf\Sigma}$ to the Dyson equation owing to
the coupling of the source and drains,
and a total self energy becomes ${\bf\Sigma}= \tilde{\bf\Sigma}+{\bf \Sigma}^{imp}$.
Here, the self-energy $\tilde{\bf\Sigma}(E)$ reads,
\begin{eqnarray}
\tilde{\bf\Sigma}_{lm:l'm'}(E) = t_H^2 \delta_{mm'}\Big[
\delta_{m,0} {\bf G}_{l(-1):l'(-1)}(E)|_{\mu=\mu_S} \nonumber\\
+\delta_{m,M-1} {\bf G}_{lM:l'M}(E)|_{\mu=\mu_D} \Big ]
\label{SigmaLead}
\end{eqnarray}
where the subscripts of $\mu=\mu_{S,D}$ denote that
each equilibrium Green's function is determined by
different chemical potentials of  $\mu_S=\mu_0-eV_S$ and $\mu_D=\mu_0-eV_D$
accounting for applied voltages, $V_S$ and $V_D$  at each reservoir, respectively.

Solutions of the Dyson equation are obtained by inverting the matrix equation Eq. (\ref{Dyson}).
Firstly, it's retarded component is calculated as,
\begin{equation}
{\bf G}^R(E)  = [({\bf g}^R)^{-1}-{\bf \Sigma}^R]^{-1}.
\label{GR}
\end{equation}
Here, ${\bf g}^R(E)=[E{\bf 1}-{\bf H}-{\bf v}]^{-1}$ is the free-particle
Green's function and 
${\bf\Sigma}^R(E)= \tilde{\bf\Sigma}^R(E)+{\bf \Sigma}^{imp,R}(E)$ is
a retarded component of the self-energy.
Detailed form of $\tilde{\bf\Sigma}^R(E)$ is given in the Appendix.
Whereas, the term of ${\bf \Sigma}^{imp,R}(E)$ depends on diagonal components
of it's own Green's function, as indicated by Eq. (\ref{ImpSigma}). 
Thus, we should solve the above matrix equation self-consistently.

With the obtained ${\bf G}^R$ and it's Hermitian conjugate ${\bf G}^A$,
the Keldysh components of the Green's function and the self-energy become
\begin{equation}
{\bf G}^K(E)  = {\bf G}^R(E) {\bf \Sigma}^K(E) {\bf G}^A(E)
\label{GK}
\end{equation}
and 
\begin{equation}
{\bf\Sigma}^K(E)= \tilde{\bf\Sigma}^K(E)+{\bf \Sigma}^{imp,K}(E),
\end{equation}
respectively.
According to Eq. (\ref{SigmaLead}), the self-energy contributed from the the source
and drain coupling is obtained as,
\begin{eqnarray}
 \tilde{\bf\Sigma}_{lm:l'm'}^K(E) = \tilde{\bf\Sigma}_{lm:l'm'}^C(E)
\Big[ \delta_{m,0} \tanh(\frac{E-\mu_S}{2 k_B T}) \nonumber\\
+\delta_{m,M-1} \tanh(\frac{E-\mu_D}{2 k_B T}) \Big ]
\label{SigmaLeadK}
\end{eqnarray}
with $\tilde{\bf\Sigma}^C(E)= \tilde{\bf\Sigma}^R(E)- \tilde{\bf\Sigma}^A(E)$,
the correlated component of the self-energy.
However, for the Keldysh component of the impurity-induced self-energy
${\bf\Sigma}^{imp,K}(E)$ the result is not given in a closed form 
and should be calculated self-consistently
as in the case of the retarded one via Eqs. (\ref{ImpSigma}) and (\ref{GK}).

\subsection{Electron density and current}

The ensemble average of $ n_{lm}=\langle b_{lm}^\dagger b_{lm}\rangle$ gives
local electron density of the device and, consequently, the electron density
distribution in Eq. (\ref{Poisson}) becomes
$n_{el}({\vec r})=\sum_{lm} n_{lm}\chi_l({\vec\rho})\psi(z_m)$.
>From the generating functional technique as in Ref. \cite{Oh0},
one can express the local electron density with the calculated Green's functions. The result reads,
\begin{eqnarray}
n_{lm}&=&  \frac{1}{2a}\Big[1-\frac{i}{2\pi}\int_{-\infty}^{\infty}dE~ {\bf G}^K_{lm:lm}(E)\Big]\nonumber\\
      &=&    {\rm tr}\int_{-\infty}^{\infty}dE~ \Big[ {\bf f}_{FD}(E) {\bf D}(lm:E) \Big].
\end{eqnarray}
Here, in the second line we use the functional form of Fermi-Dirac
distribution ${\bf f}_{FD}(E)$ and the density-of-state ${\bf D}(lm:E)$
for the resemblance with equilibrium results.
Since the device is in non-equilibrium condition, two functions are given in a matrix form;
the Fermi-Dirac distribution matrix is defined by,
\begin{eqnarray}
{\bf f}_{FD}(E) &=& \frac{1}{2} [{\bf 1}- {\bf \Sigma}^K({\bf \Sigma}^C)^{-1}]
\label{fFD}
\end{eqnarray}
while, using Eq. (\ref{GK}), the density-of-states matrix at the node $m$ and transverse mode $l$,
is expressed by,
\begin{eqnarray}
{\bf D}(lm:E) &=& \frac{ig_{sv}}{2\pi a} {\bf\Sigma}^C {\bf G}^A{\bf 1}_{lm} {\bf G}^R.
\end{eqnarray}
Here, $g_{sv}=4$ is the spin-valley degeneracy,
${\bf\Sigma}^C={\bf\Sigma}^R-{\bf\Sigma}^A$,
and ${\bf 1}_{lm}$ is a matrix whose elements
are non-zero only at the $lm$-th diagonal position.
When the impurity scattering is absent, ${\bf f}_{FD}$ becomes
the well-known results as in Ref. \cite{Wang,Datta}, where
non-zero elements are only at $m=0$ and $m=M-1$ nodes and
are equal to the Fermi-Dirac distribution characterized by $\mu_S$ and$\mu_D$,
respectively.
However,  due to the impurity scattering of ${\bf \Sigma}^{imp}$,
elements of ${\bf f}_{FD}$ are deviated from the Fermi-Dirac distribution function in general.

Currents flowing through the device is defined by time-derivatives of total charge
at nodes $m=-1$ or $m=M$.
Then, through the Heisenberg equation of motion, one can find that the currents becomes,
\begin{eqnarray}
I_{DS} &=&-\frac{e}{2\pi\hbar} {\rm tr} \Re\int_{-\infty}^{\infty} dE~
\Big[
{\bf G}^R {\bf 1}_{m}{\bf\Sigma}^K{\bf 1}_{m}\!+\!{\bf G}^K{\bf 1}_m{\bf\Sigma}^A{\bf 1}_m
\Big]\nonumber\\
&=&-\frac{e}{2\pi\hbar} {\rm tr} \int_{-\infty}^{\infty} dE~ {\bf f}_{FD}(E) {\bf T}_{m}(E)
\label{IDS}
\end{eqnarray}
where, by $m=0$ or $M-1$, the expression means currents at the source or the drain,
respectively, and ${\bf 1}_m=\sum_l {\bf 1}_{lm}$.
In the second line of the above equation, we define the transmission matrix
${\bf T}_m$ by,
\begin{eqnarray}
{\bf T}_m = g_{sv} {\bf\Sigma}^C\Big( {\bf 1}_m{\bf G}^R{\bf\Sigma}^C{\bf\Sigma}^A{\bf 1}_m
-{\bf G}^A {\bf 1}_m {\bf\Sigma}^C {\bf 1}_m {\bf\Sigma}^R\Big).
\label{Tcoeff}
\end{eqnarray}
In the case of free impurities, this form also recovers
the previous results.\cite{Wang,Datta}

\subsection{Approximations}
Prior to numerical calculations, let us first look at the approximations used.
Firstly, we consider a finite number $N$  of transverse modes.
%however,  which is sufficiently
%large to account for effects of thermally-excited electrons.
Then, the solution of Eq. (\ref{GR}) is obtained by inverting a $(NM)\times (NM)$ matrix iteratively.
However, this scheme demands the huge computational cost
because the matrix size is large and is deviated from the tridiagonal form
due to off-diagonal elements of
the self energy ${\bf\Sigma}$ and the Hamiltonian ${\bf v}$.

As an approximation, we consider leading terms in Green's functions
to emphasize mainly the effects of the impurity scattering.
This is equivalent to consider the diagonal components of the Green's functions
for transverse modes. Namely,
the coupling of different transverse modes in the self energy ${\bf\Sigma}$
and the Hamiltonian matrix ${\bf v}$ are neglected.
As indicated in Ref. \cite{Wang},
if the potential energy $U({\bf r})$ is
a slowly-varying function along the radial direction at any node $m$
the Hamiltonian matrix ${\bf v}$ becomes small and  the approximation is well justified.
As for the self-energy, leading terms in the Green's functions are obtained by writing
overlap functions of Eq. (\ref{Overlap}) as,
\begin{eqnarray}
S_{lml'm:l_1ml_2m}\simeq \delta_{l_1l_2} \frac{u_0^2l_s^3}{2a}
\langle \chi_l\!\mid\! \mid\chi_{l_1}(\rho)\mid^2 n_D(\rho,z_m)
\mid\!\chi_{l'}\rangle \nonumber\\
\simeq \delta_{l_1l_2}\delta_{ll'} \frac{u_0^2l_s^3}{2a}
\langle \chi_l\!\mid\! \mid\chi_{l_1}(\rho)\mid^2 n_D(\rho,z_m)
\mid\!\chi_{l}\rangle
\label{ImpApp}
\end{eqnarray}
and, therefore, the self-energy of Eq. (\ref{ImpSigma}) becomes diagonal for transverse modes.
However, the approximation of Eq. (\ref{ImpApp}) still couples transverse modes 
non-trivially because each diagonal component of the self-energy depends on others.

Another approximation is made in the Keldysh component of the impurity self-energy  ${\bf\Sigma}^{imp,K}$.
After various numerical calculations, we find that 
${\bf\Sigma}^{imp,K}$ is well represented by;
\begin{eqnarray}
{\bf\Sigma}_{lm:l'm'}^{imp,K}(E) = \delta_{ll'}\delta_{mm'}{\bf\Sigma}_{lm:lm}^{imp,C}(E)
~~~~~~~~~~~~~~~\nonumber\\
\left\{
\begin{array}{ll}
\tanh(\frac{E-\mu_S}{2 k_B T})&~~~~~~~~{\rm for}~~ m<    M/2 \\
\tanh(\frac{E-\mu_D}{2 k_B T})&~~~~~~~~{\rm for}~~ m\geq M/2 \\
\end{array}
\right.
\label{SigmaKApp}
\end{eqnarray}
where a node $m=M/2$ is the middle point in the intrinsic Si wire.
This indicates that particles at the nodes near the source(drain) have still the chemical potential
$\mu_S$ ($\mu_D$), not an intermediate value between $\mu_S$ and $\mu_D$,
even after suffering from scattering.
We attribute this result to a particular potential distribution in the device of 
a source-to-channel barrier, which prevents particles with different chemical potentials
from mixing.

\section{Results and Discussions}

In this section, we numerically illustrate solutions of
the non-equilibrium Green's functions suffering ionized impurity scattering
and related transport properties.
%within the self-consistent Born approximation.
We consider a typical case of the device structure which can be realized experimentally.
As shown in Fig. \ref{figSt}, the source and drain regions
are doped at $10^{20}/{\rm cm}^3$ and there is no gate-to-source and -drain overlaps to
constitute nearly abrupt junctions with the intrinsic channel.
The source and drain extensions are $15$nm and the gate length $L_G$ is $20$nm,
so that a total device length simulated is $50$nm.
By choosing a node spacing of $a=0.25$nm,  we have the number of $200$ nodes
along the wire.
In order to highlight quantum effects, we choose a small radius ($3$nm) of the wire
which exhibits $3$ mode occupancies at a zero temperature.
However, to include thermally excited particles as well as the mode coupling from the impurity scattering,
$20$ transverse states are incorporated.
The gate oxide layer has a thickness of $2.5$nm and is treated as an infinite potential barrier
for electrons. Due to this, wavefunctions at the interface between the Si wire and the oxide 
are assumed to be zero in all of our simulation.

The Poisson's equation is solved in the cylindrical coordinates
with Dirchlet boundary conditions at the gate-oxide interface,
otherwise, with Neumann conditions.
For a rapid convergence of solutions, we use the Newton-Rhapson method for
the Gummel form of external charges.\cite{Gummel}
To model a gate material, we choose a work function of $4.56eV$, approximately for TiN.
%at which
%the band diagram of the device is flat if the entire Si wire is uniformly doped at
%$10^{20}/{\rm cm}^3$.

\begin{figure}
\centering
\includegraphics[width=0.4\textwidth]{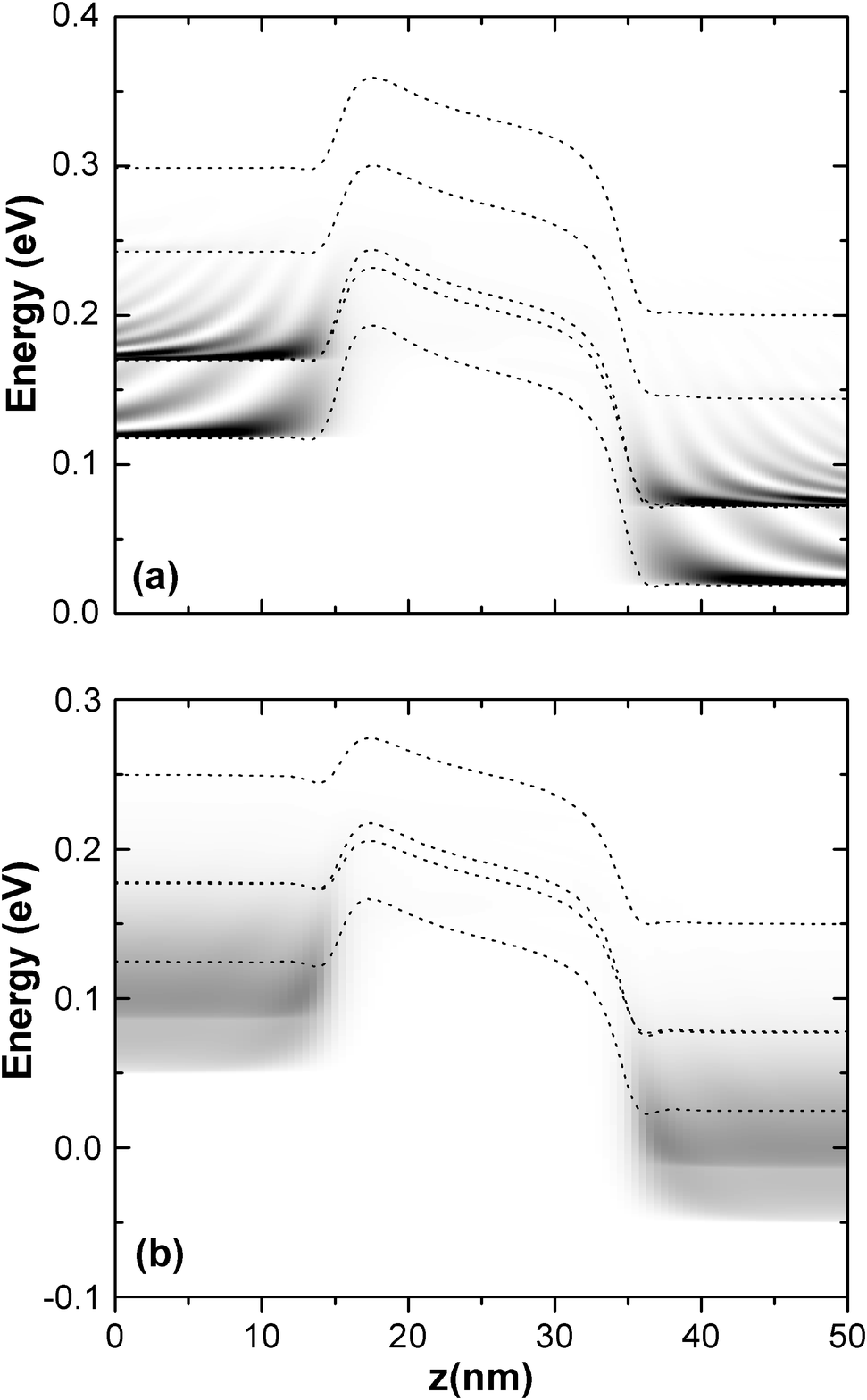}
\caption{\label{Fig2}
For a cylindrical Si wire with a 20nm gate length, we plot local particle density $\sum_l n_{lm}$
along the wire for impurity scattering strengths of $u_0=0$ and $39(eV)$ in $(a)$ and $(b)$,
respectively, at $V_{GS}= 0.6$V, $V_{D}=0.1$V, $V_S=0V$, and $T=300$K.
For qualitative comparison, we display  higher density with a darker color.
Dotted lines describes the effective potential energy of each subband before they are renormalized by
impurity scattering. 
}
\end{figure}
In Fig. \ref{Fig2}, we show calculated electronic subbands of each level
and local particle density along the wire,
and compare the results with and without the impurity scattering in $(a)$ and $(b)$,
respectively($V_G=0.6V$ and $V_D=0.1V$).
The subband bottoms(dotted lines) reflect the calculated self-consistent potentials
in which electrons at each levels feel at a node $m$.
Regardless of the impurity scattering, they exhibit 
source-channel barriers. Since a high gate voltage lowers the energy barriers,
the basic transistor action is achieved by controlling these barriers.\cite{Johnson}

The energy-resolved particle density is plotted in a gray scale; 
a darker area in the figure represents  higher density.
In the impurity-free case of $(a)$, since there is no momentum relaxation,
states injected from the drain(source) end of the device undergo reflections and
interfere strongly to the right(left) of the source-to-channel barrier.
This interference results in coherent oscillations in the particle density as seen in Fig. \ref{Fig2}-(a).
As a function of energy, it is found that the local particle density far from the source-channel barrier
shows sharp peaks like $1/\sqrt{E}$ at every onset of subbands,
reminiscence of one-dimensional density of states.

If one turns on impurity scattering,
phase information of the electrons within the device is randomized and
the energy levels are renormalized.
Above all, this makes the interference oscillations washed out
in the local particle density as shown in Fig. \ref{Fig2}-(b).
%the oscillating patterns disappear completely for
%a strong impurity scattering. 
In addition, electronic states are shifted and broaden, so that
the most electrons are found below subband bottoms and
it's occupation has no longer $1/\sqrt{E}$-dependence, but a monotonically varying function(the abrupt change
of darkness along the energy direction comes from a different valley state).
In both cases of the impurity scattering,
one can see that electrons in the source and drain regions are well separated
by the source-channel barriers from each other.
Due to this, the approximation of Eq. (\ref{SigmaKApp}) is justified with good accuracy.
%and tunneling process may contribute mainly to transport through the barriers.

\begin{figure}
\centering
\includegraphics[width=0.4\textwidth]{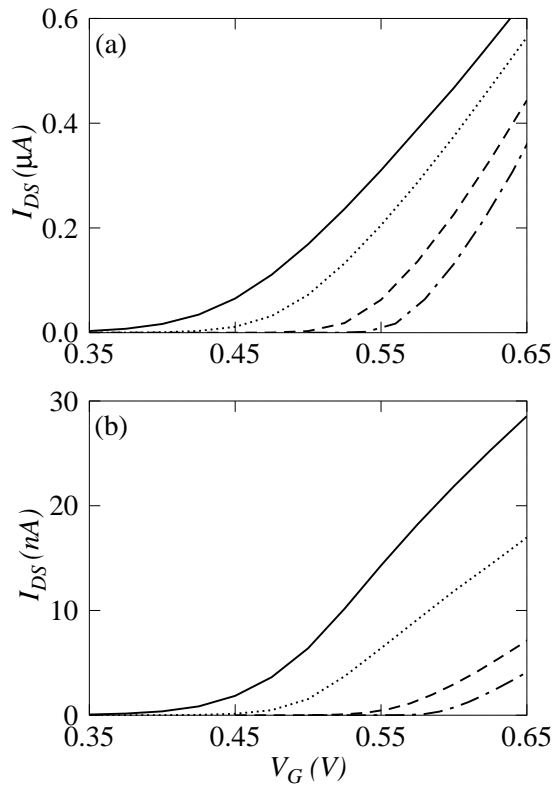}
\caption{\label{figIDVG}
We compare calculated $I_{DS}$-$V_{G}$ results at temperatures of $300$K(solid),
$200$K(dotted), $100$K(dashed), and $50$K(dot-dashed lines), respectively,
for impurity scattering strengths of $u_0=0$ in $(a)$ and $39eV$ in $(b)$.
Here, we assume a small source-drain bias of $0.02V$.
}
\end{figure}
In order to examine the electronic transport of the device, we calculate
channel currents $I_{DS}$ versus a gate voltage $V_G$
at a small source-drain bias, and plot results in Fig. \ref{figIDVG}-(a) and (b), respectively,
with and without impurity scattering for several temperatures.
Under this condition, currents exhibit rapidly increasing behavior as a gate voltage becomes larger.
This shows the basic operation of a transistor as indicated in the previous section;
the channel current turns on by lowing the source-channel barrier when a gate voltage is higher than
a certain value, called a threshold voltage $V_{th}$.

By comparing Figs. \ref{figIDVG}-(a) and (b) at a given temperature,
one can find that the presence of impurities reduces the currents significantly
even though electrons in both cases are expected to move ballistically in the intrinsic gate region.
This indicates that transport through the Si wire largely depends on the electronic structure
of the source and drain regions.

%and, thereby, potential drops occurs in their extensions as well as the intrinsic region.
%Consequently, this effectively lowers the potential difference across the intrinsic region
%and may be responsible for  the reduced currents.
As inspired by flat subbands in the figures,
the potential drops across the intrinsic regions are nearly invariant to the impurity scattering strength.
Thus, it is reasonable to assume that the suppressed currents do not come from
the Fermi-Dirac matrix of Eq. (\ref{IDS}) which crucially depends on the potential drop,
but mainly from a reduced transmission coefficient of Eq. (\ref{Tcoeff}).
One of possible explanations for this is that electrons injected from the source are partially
reflected from impurities in the source extension in addition to that from the source-channel barriers
and, thus electrons tunnel the source-channel barrier at rare intervals.
This type of the reduction for the transmission coefficient is also encountered
in problems of tunneling in dissipative environments.\cite{Oh1,Ingold}
According to the theories, when environments of the device become  more dissipative,
carriers are harder to tunnel the barriers because more energies should be transferred to
the environment.

As a function of a temperature,
curves are shifted with wholly similar shapes and
slightly different slopes in both cases of the impurity scattering.
Two points are noteworthy.
Firstly, the threshold voltage is shifted to a higher value as a temperature is lowered.
This is easily understood because as the temperature decreases, available electrons
to overcome source-to-channel barrier thermally are reduced and then
more potential energy should be supplied electrostatically to turn on currents.
Secondly, we look at the slopes of the $I_{DS}$-$V_G$ curves.
In conventional MOSFETs, they are related to a channel mobility $\mu_m$ 
via a relation of $ I_{DS}\propto \mu_m (V_G-V_{th})(V_{D}-V_{S})$.
As seen in the figures, our results show linear behavior in some range of
gate voltages. Therefore, we may understand the slopes to be proportional to the mobility of electrons in the device. 
For detailed comparison, we define the conductance by
\begin{equation}
\sigma(T) = \left.\frac{\partial I_{DS}}{\partial V_G} \right|_{V_G=0.65V},
\end{equation}
known as the transconductance in MOSFETs.

\begin{figure}
\centering
\includegraphics[width=0.4\textwidth]{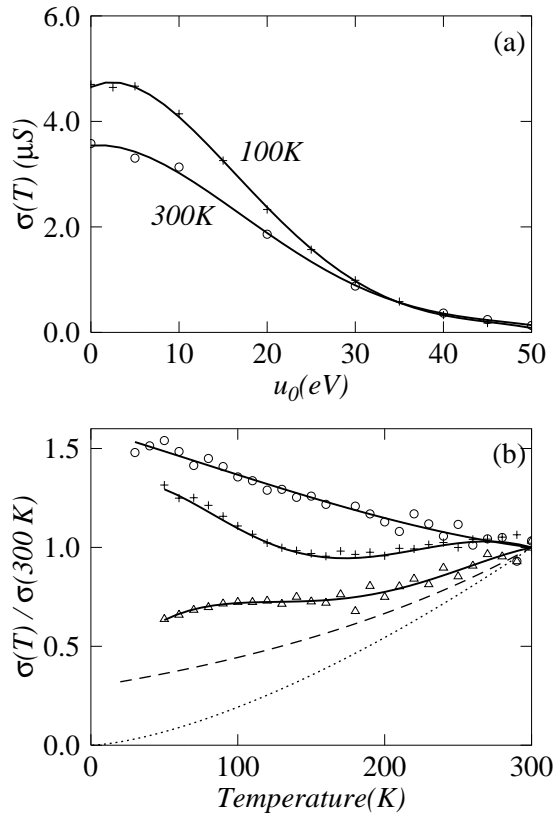}
\caption{\label{figmob}
In $(a)$, we plot calculated conductances(symbols) as a function of impurity scattering potential
at two different temperatures of $300$K and $100$K, respectively($V_{D}=0.02V)$.
In $(b)$, calculated conductances are plotted as a function of temperature
for given impurity scattering potential of $u_0=0$(circles), $23eV$(crosses),
and $39eV$(triangles), respectively.
Solid lines are just guide to the eye.  To emphasize their temperature dependence
we normalize them with values at $300$K and superimpose
the lines of $e^{1.22(T/300{\rm K}-1)}$(dashed) and 
$(T/300{\rm K})^{3/2}$(dotted).
}
\end{figure}
Calculated conductance is summarized in Fig. \ref{figmob} as functions of impurity scattering
strength and temperature.
In Fig. \ref{figmob}-(a) we compare the conductance with increasing
impurity scattering strength for two temperatures.
It is noted that the conductance decreases monotonically when the impurity scattering strength becomes larger
at both temperatures and, consequently, suppressed mobilities are expected.
In Fig. \ref{figmob}-(b) we plot the temperature dependence of the conductance
for various impurity scattering strengths.
For a bulk material, it is well known that
the mobility resulted from impurity scattering is proportional to $T^{3/2}$
to the first order(dotted line in the figure).\cite{Smith}
In the case of a two-dimensional system,
the ionized impurity scattering (for instance, in a quantum well with a $\delta$-doping) is enhanced
due to the increased overlap of the ionized impurity with electron wavefunctions and
the mobility decreases nearly exponentially
when a temperature is lowered(dashed line).\cite{Masselink}
In our case of a quasi-one-dimensional system,
the conductance shows different temperature dependences
from those of higher-dimensional ones; 
the conductance of the Si wire interpolates from 
linearly increasing behavior of the impurity-free case
to the exponentially decaying dependence of a strong impurity scattering
as a function of scattering strength.
Curves shown in Fig. \ref{figmob}-(b) do not provide a definitive comparison of ionized-impurity
scattering among three different dimensional systems because each system has different
doping profiles and concentrations.
Despite of this, it is interesting to note that the ionized impurity scattering becomes
less temperature-dependence when the system has a lower dimension.

\section{Summary}

In summary, we study transport through a gate-all-around Si wire
in the ballistic regime by considering the ionized impurity scattering.
Using the Schwinger-Keldysh approach, we include the impurity scattering
within the self-consistent Born approximation and present expressions for electron density and currents
in terms of non-equilibrium Green's functions and self-energies. 
By simulating a typical case of a Si wire, we compare electron densities and channel currents
for zero- and strong-impurity scattering strengths.
In the case of the strong impurity scattering, we find that
the local particle density profiles are shifted and broaden to
result in suppressed currents compared to the zero-impurity scattering case,
and the oscillating interference pattern vanishes.
Calculated currents and conductances are also presented as  functions of temperature
and the impurity scattering strength.
It is found that the conductance of a Si wire exhibits various behavior by decreasing temperature,
which interpolate from a linear increasing function at a zero scattering to an exponentially
decreasing function for the strong scattering case.
However, in this work, we do not include other inelastic scattering process such as acoustic and optical 
phonon scattering which will be occurred in real devices.
Therefore, our results  show the effects of
the ionized impurity scattering alone on electronic transport through a Si wire.
%In further work we will
%focus our attention on transport affected by various scattering mechanism.

\acknowledgments{
The authors would like to thank M. Shin for useful discussions.
This work was supported by the Korean Ministry of Science and Technology
through the Creative Research Initiatives Program under Contract No.
R17-2007-010-01001-0(2007).}

%\newpage

\appendix

\section{\label{GreenLead} Green's functions of a uniformly-doped wire }

In this Appendix, we illustrate Green's functions for an infinitely long Si wire which is doped uniformly.
Eigenstates are plain waves whose wavelength is determined by periodic boundary conditions.
Since the wire is translational invariant,
the self-energy in Eq. (\ref{ImpSigma}) is independent of a longitudinal position.
Then,  the retarded component of the Green's function can be derived as,
\begin{eqnarray}
{\bf G}_{lm:l'm'}^R(E) = \frac{\delta_{ll'}}{M_0} \sum_{k=-M_0/2}^{M_0/2}
\frac{e^{2\pi i k(m-m')/M_0}}{E-\epsilon_{lk} -{\bf\Sigma}^{imp,R}_{lm:l'm'}(E)}
\end{eqnarray}
where $M_0$ is the number of  a longitudinal node and $\epsilon_{lk}=\epsilon_l +2 t_H[1-cos(2\pi k/M_0)]$ 
with an eigenenergy $\epsilon_l$ of the $l-$th transverse mode.
In the limit of a large $M_0$, diagonal components of the Green's function
reads;
\begin{eqnarray}
{\bf G}_{lm:lm}^R(E) = \frac{1}{4t_H} \frac{{\rm sign}(y_l(E)-1/2)}{\sqrt{y_l^2(E)-y_l(E)}}
\label{AGreen}
\end{eqnarray}
with
\begin{eqnarray}
y_l(E) = \frac{E-\epsilon_l-{\bf\Sigma}^{imp,R}_{lm:lm}(E)}{4t_H}\nonumber
\end{eqnarray}
and, according to Eqs. (\ref{ImpSigma}) and (\ref{ImpApp}),
the self-energy is proportional to the diagonal component of the Green's functions like
\begin{eqnarray}
{\bf\Sigma}^{imp,R}_{lm:l'm'}(E) = \delta_{mm'}\delta_{ll'}\sum_{l_1}
S_{lmlm:l_1m:l_1m} {\bf G}^R_{l_1m:l_1m}.
\label{AImpApp}
\end{eqnarray}
Thus, the Green's functions are obtained by solving Eqs. (\ref{AGreen})
and (\ref{AImpApp}) self-consistently.
On the other hand, the chemical potential $\mu_0$  of the uniformly doped wire can be found
from the particle density of $n_{lm}=-g_{sv}\Im{\bf G}_{lm:lm}^R(E)/\pi a$
together with the Poisson's equation.

The self-energy of Eq. (\ref{SigmaLead})
caused by the coupling of the device to the source and drain regions
is obtained by solving the uniformly doped wire with vanishing boundary conditions.
In the similar way to Eq. (\ref{AGreen}), it is given by,
\begin{eqnarray}
\tilde{\bf\Sigma}^R_{lm:l'm}(E) &=& \delta_{ll'} t_H[2y_l(E)-1] \nonumber \\
&&\left[1-\sqrt{1-\frac{1}{[2y_l(E)-1]^2}}~~ \right].
\end{eqnarray}

\end{document}